\documentstyle[epsfig]{elsart}
\newcommand{\hhm}{\alpha}
\newcommand{\hhn}{\beta}

\newcommand{\be}{\begin{equation}}
\newcommand{\ee}{\end{equation}}
\newcommand{\bea}{\begin{eqnarray}}
\newcommand{\eea}{\end{eqnarray}}

\newcommand{\A}{{\bf a}}

\newcommand{\bbf}{\bf}
\newcommand{\ssl}{\sl}

\newcommand{\T}{\mbox{\bf T}}

\begin{document}

\begin{frontmatter}
\begin{flushright}
MPI-PHT-99/18 \\
FSUJ-TPI-99/04 \\
hep-th/9904174  \\
April 1999
\end{flushright}

\vspace{3 cm}

\title{Bogomol'nyi Equations for \\
Einstein--Yang--Mills--Dilaton theory}

\author[MSV]{Mikhail S. Volkov \thanksref{DFG}} and
\author[DM]{Dieter Maison}
\address[MSV]{Institute for Theoretical Physics\\
Friedrich Schiller University of Jena\\
Max-Wien Platz 1, D-07743, Jena, Germany.\\
e-mail: vol@tpi.uni-jena.de}
\address[DM]{Max-Planck-Institut f\"ur Physik\\
--- Werner Heisenberg Institut ---\\
F\"ohringer Ring 6, 80805 Munich, Germany.\\
e-mail: dim@mppmu.mpg.de}

\thanks[DFG]{Supported by the Deutsche
Forschungsgemeinschaft, DFG-Wi 777/4-1}

\begin{abstract}
A static, spherically symmetric and purely magnetic 
solution of the Einstein-Yang-Mills-Dilaton theory,
found previously by numerical integration 
is shown to obey a system of first order Bogomol'nyi equations. 
As common for such equations, there is a tight relation to supersymmetry,
in the present case to the N=4 gauged SU(2)$\times$SU(2) supergravity of 
Freedman and Schwarz. Specifically, the dilaton potential of the latter
can be avoided by choosing one of the two gauge coupling constants
to be imaginary. 
It is argued that this corresponds to a hitherto unknown N=4 gauged
SU(2)$\times$SU(1,1) supergravity in four Euclidean dimensions leading to 
Bogomol'nyi equations with asymptotically flat solutions. 
\end{abstract}

\end{frontmatter}
\newpage

{\bf Introduction.--} 
Supergravity backgrounds play an important role in the analysis
of string theory. Besides genuine fully supersymmetric string vacua,
also particle like solutions with partial supersymmetry
(p-branes, monopoles etc. \cite{Duff95})  are
presently obtaining much consideration, in particular in view
of their role in verifying various duality conjectures \cite{Hull98}.
However, apart from stringy monopoles
and the related solutions \cite{Duff95}
obtained via the heterotic five-brane
construction \cite{Strominger90},
most of the literature is devoted to solutions with Abelian gauge fields.
This is easily understood, since such configurations can be obtained
straightforwardly from the known solutions of the Einstein-Maxwell system.
On the other hand, it is to be expected that also configurations with
non-Abelian gauge fields will eventually play an important role.  
In addition, gauged supergravity models have recently regained 
considerable interest in view of the AdS/CFT correspondence
\cite{Maldacena97}, which also suggests studying classical solutions
of supergravities with non-Abelian gauge fields.

The incompleteness of the Abelian picture can be observed already
in the (non-supersymmetric) Einstein-Yang-Mills (EYM) theory.
A number of interesting results have been obtained after the discovery
in this theory of particle-like solutions by Bartnik and McKinnon
\cite{Bartnik88} (see \cite{Volkov98} for a recent review).
In view of the complexity of the field equations even in the case of
static, spherically symmetric solutions, our knowledge is
largely based on numerical analysis. Nevertheless many interesting and
partly surprising results are available
as well on globally regular solutions as on black holes with
`non-Abelian hair'.
In particular they show that a number of the standard electrovacuum theorems 
{\em do not apply}
in the non-Abelian domain:
\begin{itemize}
\item
The Birkhoff Theorem is not valid, i.e. there are time-dependent spherically
symmetric solutions.
\item
There exist globally regular, asymptotically flat static solutions.
\item
Static black holes are not uniquely specified by their mass and their `gauge'
charges -- the `No Hair' Conjecture is violated.
\item
Static black holes are not necessarily spherically symmetric --
Israel's theorems do not apply. 
\item
Non-rotating stationary black holes are not necessarily static --
the Abelian staticity conjecture does not apply.
\end{itemize}
In view of these results it becomes clear that experience gained
in the Abelian domain is not universal, and
the non-Abelian sector  therefore should also be studied.

We consider in this paper certain
particle-like solutions of the coupled Einstein-Yang-Mills-dilaton
(EYMD) system.
These solutions were obtained numerically some time ago
\cite{Lavrelashvili93,Donets93,Bizon93}. 
They are static, spherically symmetric, globally regular, 
asymptotically flat and neutral --  with
the purely magnetic Yang-Mills (YM)
field strength decaying as $1/r^3$ for $r\to\infty$.
In view of their instability % against decay to the vacuum
and owing to some topological considerations
these solutions may be justly
considered as a kind of `gravitational sphalerons' \cite{Volkov98}. 
There is an infinite discrete family of such solutions labeled by the 
node number $n$ of the gauge field amplitude.
Let us call these solutions EYMD solitons.
A very special role is played by the lowest non-trivial -- the $n=1$ --
solution. Its parameters show some surprising
regularities absent for all the
higher ($n>1$) solutions, leading to the hope 
\cite{Lavrelashvili93} to find this solution by analytical methods.
One may speculate that the field equations in this special case have some 
hidden symmetry. In fact, we shall argue in the following
that this hidden symmetry is supersymmetry, although the
corresponding supergravity is not the naively expected one related to
heterotic string theory, for which the gauge field has to be self-dual
in order to have partial supersymmetry \cite{Strominger90}.

In the recent work \cite{Chamseddine97,Chamseddine98} 
non-Abelian partially supersymmetric solutions were obtained
within the N=4 SU(2)$\times$SU(2) gauged supergravity theory,
also known as Freedman--Schwarz (FS) model \cite{Freedman78}.
In the bosonic sector this theory contains the gravitational field,
two non-Abelian gauge fields $A^{(1)a}_\mu$ and $A^{(2)a}_\mu$ with 
two independent gauge coupling constants $g_1$ and $g_2$,
an axion and a dilaton.
One can consistently truncate this theory by requiring that
$A^{(2)a}_\mu=0$, while $A^{(1)a}_\mu$ is purely magnetic,
in which case the axion can be set to zero too. 
As a result, the bosonic part of the action reduces 
to the action of the EYMD theory
plus a potential term for the dilaton:
\be                                                  \label{0}
{\rm U}(\phi)=-\frac18\,(g_1^2+g_2^2)\, {\rm e}^{-2\phi}.
\ee
The procedure to find the solutions applied in
\cite{Chamseddine97,Chamseddine98} was to derive the Bogomol'nyi
equations for  partially supersymmetric configurations. 
The corresponding solutions are globally regular (and globally hyperbolic),
but not asymptotically flat -- due to the presence of the dilaton
potential (\ref{0}). 

Our strategy to obtain the Bogomol'nyi  equation in the EYMD theory
is to `supersymmetrize' the latter by deriving it
from the FS model, which will allow us to apply the
same techniques
as in the FS case. Accordingly, we need to
get rid of the potential term in Eq.(\ref{0}),
and we do this by considering
the FS model for imaginary values of the gauge coupling
constant $g_2$:
\be                                                    \label{0a}
g_2=i|g_2|.
\ee
For $|g_2|=g_1$ the potential vanishes, and we recover the EYMD theory.
The main point is that, apart from the correct EYMD Lagrangian,
we obtain in this way, at least formally, also
the rules for computing fermionic supersymmetry variations.
Setting the latter to zero gives us the equations for  the
supersymmetry Killing spinors,
whose consistency conditions can be formulated as a set of first order
equations for the background fields, and these are compatible
with the second order field equations. 
The $n=1$ EYMD soliton fulfils
these Bogomol'nyi equations and possesses two unbroken supersymmetries
thus showing that our formal procedure makes sense,
although to our disappointment we are still not able to write down the
solution in a simple analytical form.

As a result, we add one more member to the (small) family of known
supersymmetric solutions for gravitating non-Abelian gauge fields.
As is usual in this family,
the new solution exhibits some surprising features. First,
it is quite bizarre that such a solution arises in a theory without
a dilaton potential, the latter being generically present in all
gauged supergravity models. Besides, one can wonder as to whether
the solution is supersymmetric at all, since the `imaginary trick'
employed to obtain it is  a rather formal operation.
On top of all, a certain puzzle arises, for if the solution is 
supersymmetric then one needs to reconcile
this with the fact that it is unstable. 
The answer to all these questions we shall offer is that there is in fact
a new N=4 gauged supergravity with gauge group SU(2)$\times$SU(1,1) and
vanishing dilaton potential in a 4d {\em Euclidean} space.  
%%%%%%%%%%%%%%%%%%%%%%%%%%%%%%%%%%%%%%%%%%%%%%%%%%%%%%%%%%%%%%%

{\bf The EYMD model.--} Consider the  EYM-Dilaton system
specified by the action
\be                                                       \label{1}
S=
\int \left( -\frac{1}{4}\,R+\frac{1}{2}\,\partial _{\mu }\phi \,\partial
^{\mu }\phi -\frac{1}{4}\,e^{2\phi }\,F_{\mu \nu }^{a}F^{a\mu \nu }
\right) \sqrt{-\bf{g}}\,d^{4}x.  
\ee
Here the gauge field tensor is given by 
$F_{\mu \nu }^{a}=\partial _{\mu }A_{\nu }^{a}
-\partial _{\nu }A_{\mu}^{a}+\varepsilon _{abc}A_{\mu }^{b}A_{\nu }^{c}$, and
the gauge field is $A\equiv \T_a A_{\mu }^{a}dx^{\mu}$, where
the SU(2) group generators
$\T_a$ obey $[\T_a,\T_b]=i\varepsilon_{abc}\T_c$.

Let us consider the reduction to static, spherically
symmetric configurations. We choose the 4-metric as
\be                                          \label{2}
ds^2={\rm e}^{2V}\,dt^2-{\rm e}^{2\lambda}\,d\tau^2-
{\rm e}^{2\mu}\,(d\theta^2+\sin^2\theta
\,d\varphi ^{2}),
\ee
where a gauge condition can be imposed on 
the functions $V$, $\lambda$, and $\mu$.
For the Yang-Mills field we make the usual purely magnetic ansatz
\be                                                     \label{3}
A=w\ (-\T_2\,d\theta +\T_1\,
\sin \theta \,d\varphi )+\T_3\,\cos \theta \,d\varphi .
\ee
All the functions  $V$, $\lambda$, $\mu$,
$w$ as well as the dilaton $\phi $
depend only on the radial coordinate $\tau$.
Varying the action before fixing the gauge
in the line element, we obtain the complete set of the field
equations
\bea                                                          
{\rm e}^{2\mu}\,(\mu'^2+2\mu'V'-\phi'^2)&=&2{\rm e}^{2\phi}\,w'^2+
{\rm e}^{2\lambda}\left(1-{\rm e}^{2\phi-2\mu}\,
(w^2-1)^2\right)\, ,                                     \label{a4}      \\
\left({\rm e}^{V-\lambda+2\mu}\,\mu'\right)^\prime&=&        
{\rm e}^{V+\lambda}\left(1-{\rm e}^{2\phi-2\mu}\,
(w^2-1)^2\right) \, ,                                    \label{a5}      \\
\left({\rm e}^{V-\lambda+2\mu}\,\phi'\right)^\prime&=&
2{\rm e}^{2\phi+V-\lambda}\,w'^2            
+{\rm e}^{2\phi+V+\lambda-2\mu}\,(w^2-1)^2 \, ,           \label{a8} \\
\left({\rm e}^{2\phi+V-\lambda}\, w'\right)^\prime&=&
{\rm e}^{2\phi+V+\lambda-2\mu}\, w(w^2-1) \, ,            \label{a9}  \\
\left({\rm e}^{V-\lambda+2\mu}\, (V'-\phi')\right)^\prime&=&0\, ; \label{a6} 
\eea
(all through this paper $^\prime :=\frac{d}{d\tau}$).
We note that the last equation
can be regarded as a Noether conservation law implied by the
invariance of the action under global dilatations
\be                                                    \label{a9a}
V\rightarrow V+\epsilon,\ \
\lambda\rightarrow \lambda-\epsilon,\ \
\mu\rightarrow \mu-\epsilon,\ \
\phi\rightarrow \phi-\epsilon,\ \
w\rightarrow w.\ \
\ee
As a consequence, for globally regular
solutions the following condition can be imposed:
\be                                                           \label{a10}
V=\phi-\phi_\infty
\ee
with $\phi_\infty\equiv\phi(\infty)$.
For many purposes it is convenient to implement the Schwarzshild gauge,
$\mu=\tau$, such that the geometrical radius $r={\rm e}^\tau$.
Introducing the notation $\nu\equiv{\rm e}^{\tau-\lambda}$ and
using Eq.(\ref{a10}), the metric reads
\be                                                      \label{a11}
ds^{2}={\rm e}^{2(\phi-\phi_\infty)}\, dt^{2}-{\rm e}^{2\tau}
\left(\frac{d\tau^{2}}{\nu^2}+d\Omega^2\right).
\ee
It is convenient to introduce the new variable $\psi=\phi-\tau$.
As a result, the independent field equations are obtained from
(\ref{a8}), (\ref{a9}):
\bea                                                     \label{a12}
\nu^2\psi''&+&\left(1-{\rm e}^{2\psi}(w^2-1)^2\right)\psi'+1=
2\nu^2{\rm e}^{2\psi}w'^2+2{\rm e}^{2\psi}(w^2-1)^2\, ,   \\
\nu^2 w''&+&\left(1-{\rm e}^{2\psi}(w^2-1)^2+2\nu^2\psi'\right) w'=
w(w^2-1)\, ,                                                   \label{a13}
\eea
where $\nu$ is specified by the constraint equation
\be                                                      \label{a14}
\nu^2=\frac{1-{\rm e}^{2\psi}(w^2-1)^2}
{2-\psi'^2-2{\rm e}^{2\psi}w'^2}\, ,
\ee
which is given by  Eq.(\ref{a4}), while
Eq.(\ref{a5}) is a consequence
of these equations.  We note that the dilatational symmetry (\ref{a9a})
now reduces simply to the invariance under $\tau\to\tau+\epsilon$,
which is due to the
autonomy of Eqs.(\ref{a12},\ref{a13}).
%%%%%%%%%%%%%%%%%%%%%%%%%%%%%%%%%%%%%%%%%%%%%%%%%%%%%%%%%%%%%%%%%%%%%%%

{\bf Numerical results.--}
The numerical analysis of Eqs.(\ref{a12})--(\ref{a14})
shows the existence of
an infinite sequence of asymptotically flat and globally regular solutions;
we call them EYMD solitons.
They exist in the whole interval $\tau\in(-\infty,\infty)$ corresponding to 
$r={\rm e}^\tau\in[0,\infty)$ and 
are parameterized by an integer $n=1,2,\ldots\ $, the number of zeros of the
YM-potential $w$.
Regular solutions of Eqs.(\ref{a12}-\ref{a14}) have to tend to fixed points of
these equations for $\tau\to\pm\infty$. The relevant fixed point for
solutions with a regular origin is given by
$w=\pm 1$,  $w'=0$, $\psi\equiv\phi-\tau=+\infty$, $\psi'=1$. 
Actually we may use the symmetry $w\to -w$ to select $w=1$ at $r=0$.
On the other hand the asymptotic behaviour at infinity is determined by the
fixed point $w=\pm 1$, $w'=0$, $\psi=-\infty$,
$\psi'=1$. In both limits one has $\nu=1$.

%%%%%%%%%%%%%%%%%%%%%%%%%%%%%%%%%%%%%
\begin{figure}
\hbox to\hsize{%\hss
  \epsfig{file=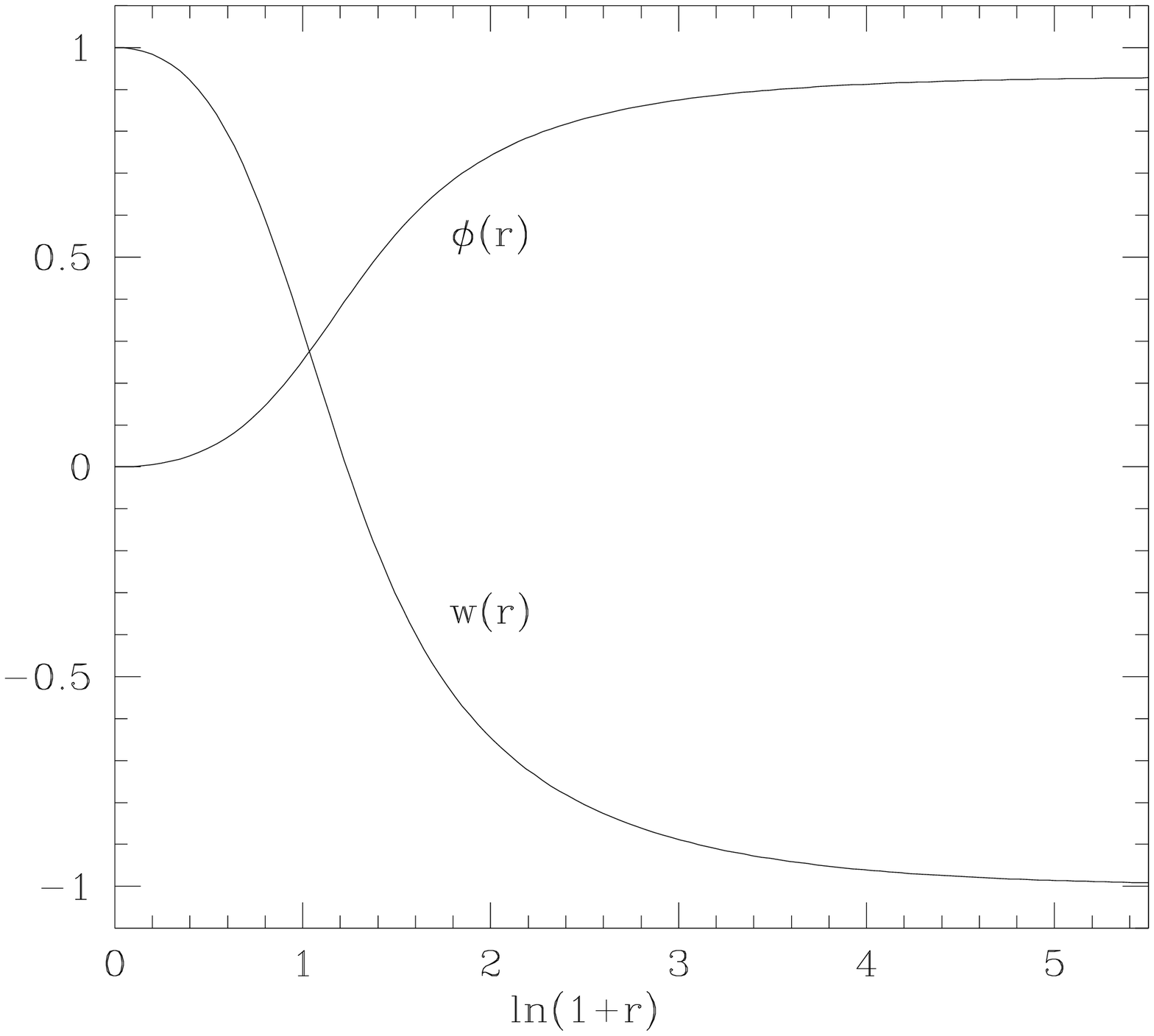,width=0.48\hsize,%
      bbllx=0.5cm,bblly=5cm,bburx=20cm,bbury=22cm}\hss
  \epsfig{file=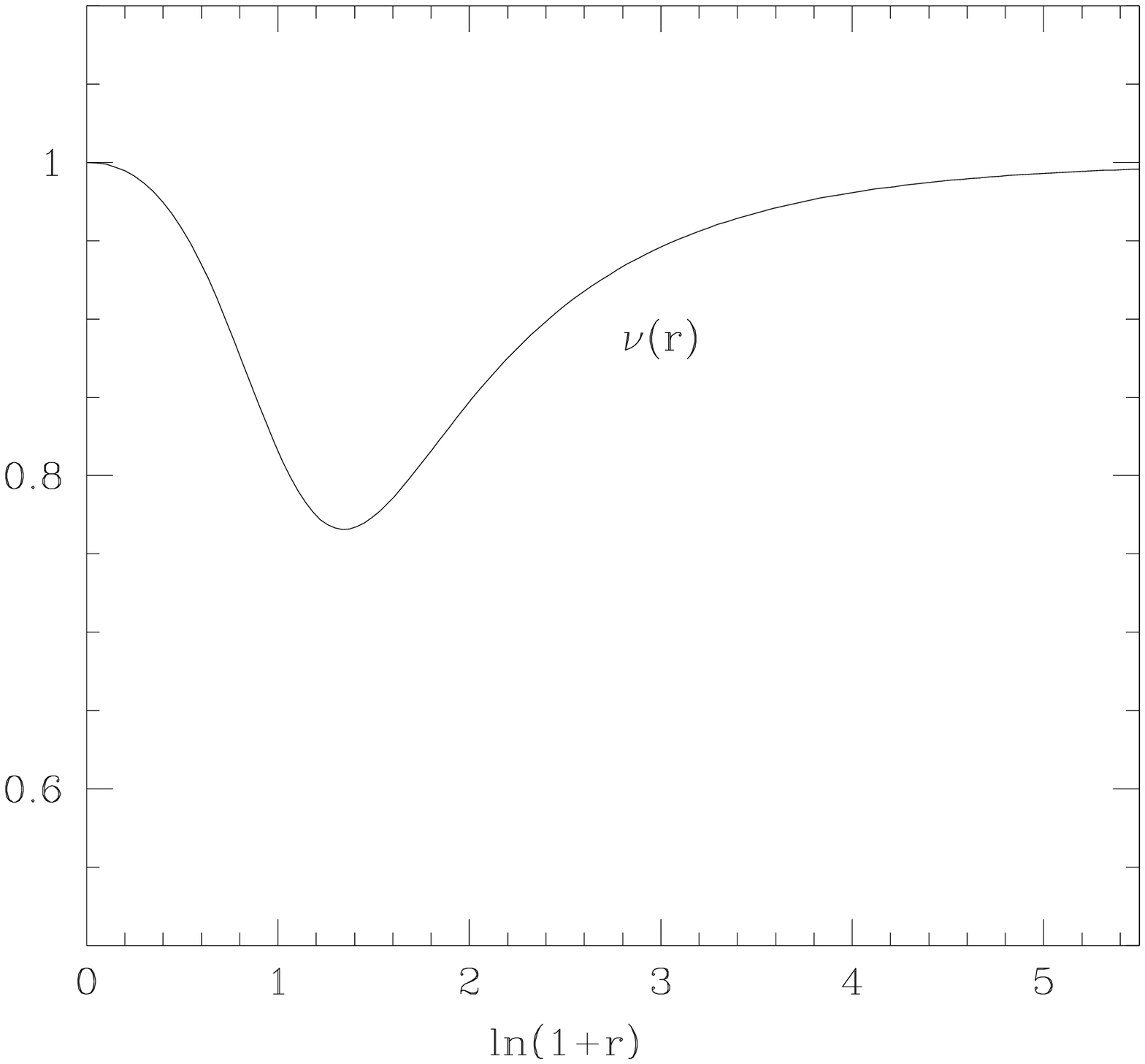,width=0.48\hsize,%
      bbllx=0.5cm,bblly=5cm,bburx=20cm,bbury=22cm}%\hss
  }
\caption{The functions $w(r)$, $\phi(r)$ and $\nu(r)$ of the 
$n=1$ EYMD soliton solution.}
\label{fig1}
\vspace{3 mm}
\end{figure}
%%%%%%%%%%%%%%%%%%%%%%%%%%%%%%%%%%%%%

Since these are `hyperbolic' fixed points some further   
relations between the otherwise independent functions $w$ and $\psi$ are
required -- they have to stay on the `stable manifold' of these fixed
points. Near the origin, $\tau\to-\infty$, this implies
\be                                              \label{6}
w=1-b\,{\rm e}^{2\tau}+O({\rm e}^{4\tau}),
\qquad\psi\equiv\phi-\tau=-\tau+\phi_0+2b^2{\rm e}^{2\tau}
+O({\rm e}^{4\tau}),
\ee
with some arbitrary parameters $b$ and $\phi_0$;
one can use the dilatational invariance to set $\phi_0=0$.
At infinity, $\tau\to+\infty$, one gets
\be                                             \label{7}
w=\pm\left(1-c\,{\rm e}^{-\tau}\right)+O({\rm e}^{-2\tau}),
\qquad
\psi=-\tau+\phi_\infty-M{\rm e}^{-\tau}
+O({\rm e}^{-2\tau}),
\ee
with arbitrary $c$, $\phi_\infty$, and $M$.
The `dilaton charge' $M$ in our case coincides with the ADM mass.
From the local point of view all the parameters
$b$, $\phi_\infty$, $c$, and $M$
are arbitrary, but for globally regular solutions 
they are fixed completely specifying the node number $n$. 

In general the numerically determined parameters of the globally 
regular solutions show no particular regularities apart from the $n=1$
solution \cite{Lavrelashvili93}. With great numerical precision one finds
\be                                             \label{10}
b=\frac16\qquad {\rm and}\qquad c=2M.
\ee
As already pointed out in the introduction, this suggests that there is
some hidden symmetry in the system, for which 
the $n=1$ solution (see Fig.\ref{fig1}) plays a special role.
%%%%%%%%%%%%%%%%%%%%%%%%%%%%%%%%%%%%%%%%%%%%%%%%%%%%%%%%%%%%%%%%%%%%

{\bf Supersymmetry.--}
We consider the bosonic part of the FS action \cite{Freedman78}:
\bea\label{11}
S_{FS}&=&\int \Biggl( -\frac{1}{4}\,R
+\frac{1}{2}\,\partial _{\mu }\phi \,\partial^{\mu }\phi
+\frac{1}{2}\, e^{-4\phi}\,\partial _{\mu }\A \,\partial^{\mu }\A
-\frac{1}{4}\,e^{2\phi }\, \sum_{\lambda=1}^2
{F}_{\mu \nu }^{(\lambda)a}F^{(\lambda) a\mu \nu }\nonumber\\
\qquad&&-\frac{1}{2}\,\A\, \sum_{\lambda=1}^2
{F}_{\mu \nu }^{(\lambda) a}\ast\! F^{(\lambda) a\mu \nu }
+\frac{1}{8}\,(g_1^2+g_2^2)\, e^{-2\phi }\Biggr) \sqrt{-\bf{g}}\,d^{4}x.
\eea
Here $F_{\mu \nu }^{(\lambda)a}=
\partial _{\mu }A_{\nu }^{(\lambda) a} -\partial _{\nu }A_{\mu}^{(\lambda) a}
+g_\lambda\,
\varepsilon _{abc}\, A_{\mu }^{(\lambda) b}A_{\nu }^{(\lambda) c}$
(there is no summation over $\lambda=1,2$),
and $\ast\! F^{(\lambda) a}_{\mu\nu}$ is the dual tensor.
One can consistently set $A_{\mu }^{(2)a}=0$.
If the field $A_{\mu }^{(1)a}$ is purely magnetic then
the invariant $\ast\! F F$ vanishes, and the axion $\A$ can be
set to zero too. With a suitable rescaling one can achieve
the condition $g_1=1$. Denoting $A^{(1)a}_\mu$ by $A^a_\mu$ 
and $g_2$ by $q$, the FS action reduces to
\be                                                    \label{16}
S_{FS}=S+\int \frac{1+q^2}{8}\,{\rm e}^{-2\phi}\sqrt{-\bf{g}}\,d^{4}x,
\ee
where $S$ is the EYMD action (\ref{1}).

In the fermionic sector the FS model contains 
four Majorana spin-3/2 fields
$\psi_{\mu }\equiv \psi _{\mu }^{\rm{I}}$ $(\rm{I}=1,\ldots 4)$ and
four spin-1/2 fields $\chi \equiv \chi ^{\rm{I}}$.
One can consistently set the fermions to zero, however,
this does not imply that their supersymmetry variations vanish:
\be                                      \label{17}
\left.
\delta \bar{\chi}=
-\frac{i}{\sqrt{2}}\,\bar{\epsilon}\,
\gamma^\mu\partial _{\mu }\phi
-\frac{1}{4}\,e^{\phi }\,\bar{\epsilon}\,
{ F}_{\mu \nu }\,\gamma^\mu\gamma^\nu+\frac{1}{4}\,e^{-\phi }\,
\bar{\epsilon}\, (1+iq\gamma_5),\right.
\ee
\be                                  \label{18}
\delta \bar{\psi}_{\rho }=\bar{\epsilon}
\overleftarrow{D}_{\rho }
-\frac{i}{4\sqrt{2}}\, e^{\phi }\,\bar{\epsilon}%
\,{ F}_{\mu\nu }\,\gamma_{\rho }\,
\gamma^\mu\gamma^\nu
+\frac{i}{4 \sqrt{2}}\, e^{-\phi }\,\bar{\epsilon}\,
(1+iq\gamma_5)\, \gamma _{\rho },
\ee
where
\be                                           \label{19}
\bar{\epsilon}\overleftarrow{D}_{\rho}\equiv
\bar{\epsilon}\left( \overleftarrow{\partial}_{\rho }
-\frac{1}{4}\,\omega ^{\ \hhm\hhn}_{\rho}\,\gamma_\hhm\gamma_\hhn
+\frac{1}{2}\,\tilde{\T}_{a}\, A_{\rho}^{a} \right),\ \ \                      
{F}_{\mu\nu}=\tilde{\T}_{a}\, F^{a}_{\mu\nu}.
\ee
Here $\bar{\epsilon} \equiv \bar{\epsilon} ^{\rm{I}}$ are spinor
parameters of the supersymmetry transformations,
$\omega ^{ \ \hhm\hhn}_{\rho}$
is the spin-connection,  and $\tilde{\T_{a}}$ 
are generators of the SU(2) subgroup of SU(2)$\times$SU(2).

Thus a bosonic configuration is invariant under 
supersymmetry 
transformations if one can find non-trivial $\bar{\epsilon}$'s
such that $\delta \bar{\chi}=\delta \bar{\psi}_{\rho }=0$.
Setting the left-hand sides of (\ref{17}) and (\ref{18})
to zero, the invariance condition becomes 
a system of linear equations for the $\bar{\epsilon}$'s.
These equations are called supersymmetry constraints and their
solutions are supersymmetry Killing spinors. 
Generically the constraint equations are inconsistent.
However, one can analyze the consistency conditions
under which non-trivial solutions for  the $\bar{\epsilon}$'s
exist. These conditions can be given in the form of a set
of nonlinear first order differential equations for the
underlying bosonic configuration -- usually called Bogomol'nyi equations.
Due to supersymmetry they are automatically compatible with the second 
order equations of motion following from the action (\ref{16}).  

The supersymmetry constraints in the model 
(\ref{16})--(\ref{19}) were analyzed in \cite{Chamseddine97,Chamseddine98}. 
It was found that if the boson fields are chosen according
to Eqs.(\ref{2},\ref{3}) then non-trivial solutions of the consistency
conditions exist, provided that the second
gauge coupling constant $q$ vanishes. 
These Bogomol'nyi equations turned out to be integrable in closed form.
Their solution describes a geodesically complete
and globally hyperbolic spacetime with
a BPS monopole type YM field (which is somewhat mysterious
since there is no Higgs field) and four independent
supersymmetry Killing spinors. 
However, due to the dilaton potential in (\ref{16}), this solution is not
asymptotically flat. 

In order to get rid of the potential in the action (\ref{16}),
let us consider the replacement
\be                                             \label{20}
q\to is
\ee
with real $s$. Obviously, for $s=1$ the potential vanishes --
the situation we
are aiming at.  At the same time, setting in (\ref{17}) and (\ref{18})
$q=is$ gives us non-trivial supersymmetry
 variations for the fermions.
It remains unclear for the time being whether such an `analytic
continuation' of the supersymmetry transformations is a legitimate
procedure, and whether the resulting transformations indeed belong to
a supersymmetry algebra. However, we temporarily ignore the problem
and go on setting the resulting variations
$\delta \bar{\chi}$  and $\delta \bar{\psi}_{\rho }$ to zero
in order to obtain the supersymmetry constraints.
 The analysis of the latter 
proceeds essentially along the same lines
as in the case for real $q$ \cite{Chamseddine98}.
One finds that in the static, spherically symmetric, purely magnetic case,
when the boson fields are chosen according to
 (\ref{2}) and (\ref{3}),
the supersymmetry constraints with $q=is$ can be made
consistent for any real $s$. The corresponding consistency
conditions are given by the relations
\bea
&&D^2\nu^2-2(B-1)^2=D^2w^2-2s^2,         \nonumber  \\
&&(B+1)^2-A^2=(s\mp C)^2,               \nonumber  \\
&&(A+B+1)(D\nu+\sqrt{2}(B-1))
+(s\mp C)(\sqrt{2}s\mp Dw)=0.      \label{21}
\eea
Here the following abbreviations are introduced:
\be                                                          \label{22}
A=2\sqrt{2}\,\nu{\rm e}^\psi(\psi'+1),\ \
B=2{\rm e}^{2\psi}(w^2-1),\ \ C=4\nu{\rm e}^{2\psi}w',\ \
D=4{\rm e}^\psi.
\ee
Under these conditions and for $s\neq 0$ there exist {\em two}
independent supersymmetry Killing spinors.
For $s=0$ the number of supersymmetries doubles,
and Eqs.(\ref{21}) reduce to the Bogomol'nyi equations studied
in \cite{Chamseddine97,Chamseddine98}.
With some labour one can verify that for any $s$
equations (\ref{21}) are compatible
with the second order field equations for the action (\ref{16}).
For $s\neq 0$ one may without loss of generality assume that $s>0$.
One can also choose the upper sign in (\ref{21}), since the other option
is recovered by replacing $w\to-w$. 

We are interested in the case where $s=1$, since then the dilaton potential
vanishes and the theory (\ref{16}) reduces to our EYMD model (\ref{1}).
The supersymmetry constraints (\ref{21}) in this case
can be expressed in the form
\bea
&&2\nu^2=w^2+1+{\rm e}^{2\psi}(w^2-1)^2, \label{23} \\
&&\nu\psi'=-\nu+(B+1)\frac{w\,(B-1)-\nu}
{B-1-8\,w\nu\,{\rm e}^{2\psi}}  \, ,                    \label{24} \\
&&\nu w'=-2\nu^2(\psi'+2)+w^2+1  \label{25}
\eea
with $B$ from (\ref{22}).
One can verify that these Bogomol'nyi equations are compatible
with Eqs.(\ref{a12})--(\ref{a14}).
Studying the power-series solutions to these equations in the
vicinity of the origin and at infinity, one immediately recovers
the relations in (\ref{10}). As a result, all hidden symmetries
are now identified and the last remaining question is whether
we can actually solve  equations (\ref{23})--(\ref{25}).
%%%%%%%%%%%%%%%%%%%%%%%%%%%%%%%%%%%%%%%%%%%%%%%%%%%%%%%%%%%%%%%

{\bf The Bogomol'nyi equations.--} 
Eqs.(\ref{23})--(\ref{25}) look rather complicated. 
However,  remarkable simplifications are possible.
With the help of Eq.(\ref{23}) $\psi$ on the right hand sides of
Eqs.(\ref{24}) and (\ref{25})
can be expressed in terms of $\nu^2$ and
$w^2$, such that the equations  assume the form
\be                                                 \label{26}
\psi'=\frac{Q_5(\nu)}{Q_2(\nu)},\ \ \
w'=\frac{R_5(\nu)}{R_2(\nu)},\ \ \
\ee
where $Q_k(\nu)$ and $R_k(\nu)$ are $k$-th order polynomials in $\nu$
with coefficients depending on $w$. Surprisingly, each pair
of polynomials in the ratios in
(\ref{26}) turns out to have a common root thus
leading to some cancellations. 
Next, one expresses $\psi'$ in (\ref{26}) in terms of $w^2$ and $\nu^2$
and their derivatives, and after this
$\nu$ can be expressed in terms of a new function $U$ via $\nu=(U+w)/2$.
As a result, Eqs.(\ref{23})--(\ref{25}) reduce to 
\bea
&&\nu^2=\frac14\, (U+w)^2 \ ,    \label{27} \\
&&\xi\,\frac{dU}{d\xi}=3-2wU-U^2 \label{28} \ ,\\
&&\xi\,\frac{dw}{d\xi}=\frac{(w^2-1)(1-wU)}{U^2-1} \ .   \label{29}
\eea
Here the new coordinate $\xi$ is related to the old one, $\tau$,
via $d\xi/\xi=d\tau/\nu$, such that the metric assumes the form
\be                                                      \label{29a}
ds^{2}={\rm e}^{2(\phi-\phi_\infty)}\, dt^{2}-{\rm e}^{2\tau}
\left(\frac{d\xi^{2}}{\xi^2}+d\Omega^2\right),
\ee
where $\phi$ and $\tau$ are now functions of $\xi$. We note that
$\phi=\psi+\tau$ and $\psi$ is expressed via $U$ and $w$
 by comparing
Eqs.(\ref{23}) and (\ref{27}). 

Now, we have two independent equations, these are  Eqs.(\ref{28}) and
(\ref{29}), and taking
their ratio the problem reduces to one non-autonomous equation
\be                                                    \label{30}
\frac{dU}{dw}=\frac{(U^2-1)(U^2+2Uw-3)}{(w^2-1)(Uw-1)} \ .
\ee
This ODE can be transformed to an Abel equation of the first kind,
but unfortunately not of a soluble type.
The boundary conditions (\ref{6}) and (\ref{7}) imply that as
$w$ varies from one (the origin) to minus one (infinity),
$U(w)$ increases from $U(1)=1$ to $U(-1)=3$.

Remarkably, Eqs.(\ref{28},\ref{29}) admit the integrable combination
\be                                            \label{35}
\frac{2\,dw}{w^2-1}-\frac{dU}{U^2-1}=\frac{d\xi}{\xi}\, ,
\ee
yielding
\be                                            \label{36}
\frac{1-w}{1+w}\,\sqrt{\frac{U+1}{U-1}}=\xi \, ,
\ee
and this allows us to separate the equations in (\ref{28},\ref{29}) to
\be                                              \label{37}
\frac{1}{2\xi}\,\frac{dw}{d\xi}=\frac{1-w^2}{4\xi^2}
-\frac{(w+1)^3}{8}+\frac{(w-1)^3}{8\xi^4}\ .
\ee
This equation admits the
discrete symmetry
\be                                               \label{38}
\xi\to\frac{1}{\xi},\ \ \ \ w\to-w.
\ee
It is worth noting in this connection that
$\xi\to 1/\xi$ is a symmetry of the line element (\ref{29a}).
We note
also that (\ref{37}) is an Abel equation of the first kind, but
unfortunately again not of any soluble type. For the regular
soliton configuration we are interested in
the asymptotic expansions of the solution are
\bea
&&w=1-\frac23\,\xi^2+\frac{2\cdot 23}{3^3\cdot 5}\,\xi^4-
\frac{2\cdot 13}{3^3\cdot 5}\,\xi^6
+\frac{2\cdot 9337}{3^8\cdot 5^2}\,\xi^8+\ldots \, \label{39}, \\
&&w=-1+\frac{2\sqrt{2}}{\xi}-\frac{2}{\xi^2}
+\frac{2\sqrt{2}}{\xi^3}-\frac{2}{\xi^4}
+\frac{{\cal C}}{\xi^5}+\ldots\ , \label{40}
\eea
where the value of the free parameter 
${\cal C}$ should be suitably chosen: ${\cal C}={\cal C}_{reg}$.
Comparing (\ref{40}) with the expansion in
Eq.(\ref{7}) one finds
${\cal C}_{reg}=2\sqrt2\,(1-4{\rm e}^{2\phi_{\infty}}/M^2)\approx -31.15$.
The numerical analysis confirms that the global solution $w(\xi)$
in the interval $\xi\in [0,\infty)$ with such asymptotics exists.
This solution is related to
the one for $w(r)$ shown in Fig.\ref{fig1} via a finite (position
dependent) rescaling of the argument.

In the generic case the asymptotic solution
for large $\xi$ is given by (\ref{40})
with ${\cal C}\neq {\cal C}_{reg}$,
while the generic regular solution near $\xi=0$
is  obtained from (\ref{40})
by applying the symmetry transformation (\ref{38}):
\be
w=1-2\sqrt{2}\xi+2\xi^2
-2\sqrt{2}\xi^3+2\xi^4-{\cal B}\xi^5+\ldots\ ,  \label{41}
\ee
where ${\cal B}$ is another free parameter.
Such solutions lead to a singular four-geometry.

Note that the emergence of the strange prime numbers in the expansion
at the origin 
in (\ref{39}) makes it appear unlikely that the analytical solution can
be represented as $w(\xi)$ in a simple closed form. 
The same problem arises if one
tries to look for the solution as $w(r)$, or $w(U)$, etc.
The strange prime
numbers appear in general also in the higher order terms
in the expansion at infinity in (\ref{40}).
However, for one particular choice, ${\cal C}=2\sqrt{2}$, all of them cancel,
and one obtains the simple solution
\be                                                 \label{42}    
w=\frac{1-2\sqrt{2}\xi+\xi^2}{1-\xi^2}\ .
\ee
Notice that this is invariant under (\ref{38}).
Unfortunately, apart from the fact that
this is not the solution we are looking for, this
solution is unacceptable, since it leads to
e$^{2\phi}\equiv 0$.

To recapitulate, assuming that
we find the solution $w\equiv w(\xi)$ of the
first order Bogomol'nyi equation (\ref{37}) with the asymptotic
behaviour (\ref{39},\ref{40}),
the full configuration is reconstructed 
as follows. One computes
\be                                                  \label{43}
U(\xi)=\frac{\xi^2(1+w)^2+(1-w)^2}{\xi^2(1+w)^2-(1-w)^2}\ ,
\ee
and
\be                                                   \label{44}
\tau(\xi)=
\ln(2\xi)+\int_{0}^{\xi}\left(\frac{U+w}{2}-1\right)\frac{d\xi}{\xi},
\ee
where the additive constant is chosen such that
the dilaton 
\be                                                  \label{45}
\phi(\xi)=
\tau(\xi)+\frac12\,\ln\left(\frac{(U+w)^2-2w^2-2}{2\,(w^2-1)^2}\right)
\ee
vanishes at the origin. The spacetime metric is given by
Eq.(\ref{29a})
whereas the gauge field is obtained from  Eq.(\ref{3}).
There are two supersymmetry Killing spinors for this configuration.
The solution describes a localized globally regular object --
a supersymmetric `gravitational sphaleron'. 
By numerical integration one finds its mass to be   
\be
M\approx 1.4657 \, ,
\ee
while the value of the dilaton at infinity
$\phi_\infty\approx 0.9322$.

{\bf Concluding remarks.--}
Employing a simple `imaginary trick' we were able to modify the Bogomol'nyi
equations of \cite{Chamseddine97} to cover the case of the EYMD 
theory, which has no 
dilaton potential. The deeper reason why this trick works is the  
existence of a hitherto unknown {\em Euclidean} version of the FS model
with gauge group SU(2)$\times$SU(1,1), whose special truncation
coincides with the static and purely magnetic sector of the EYMD theory.
The details of this new FS model will be given in
a forthcoming publication \cite{Volkov99}. In particular it will be shown
that this new N=4 supergravity
can be embedded into the standard N=1 supergravity in ten dimensions
using the SU(2)$\times$SU(1,1) group manifold as an internal space. 
Taking the negative of the Killing metric of the SU(1,1) factor for the
embedding, the contributions to the dilaton potential arising from the 
two factors cancel out, while the metric of the 4-space becomes Euclidean.

When the SU(1,1) gauge field of the new supergravity theory
is truncated, while the SU(2) one is purely magnetic and the whole
configuration is assumed to be independent of the Euclidean time,
the bosonic field equations reduce to those
of the EYMD theory in the static and purely magnetic case.
At the same time the fermionic supersymmetry
transformations exactly coincide with those obtained above via the
`imaginary trick'. As a result, the described above solution of the
EYMD theory can be viewed as a solution of the new
supergravity model, within which it becomes truly supersymmetric.
We note that
there is no paradox with the instability of the solution, since
the instability occurs
within the Lorenzian EYMD theory, which is not supersymmetric
apart from its static and purely magnetic sector.

%\bibliography{j}

\begin{thebibliography}{10}

\bibitem{Bartnik88}
R.~Bartnik and J.~McKinnon.
\newblock {\ssl Particlelike solutions of the Einstein -- Yang -- Mills
  equations}.
\newblock {\em Phys.Rev.Lett.}, {\bbf 61},\ 141--144, 1988.

\bibitem{Bizon93}
P.~Bizon.
\newblock {\ssl Saddle-points of stringy action}.
\newblock {\em Acta Phys.Polon.}, {\bbf B 24},\ 1209--1220, 1993.

\bibitem{Chamseddine97}
A.H. Chamseddine and M.S. Volkov.
\newblock {\ssl Non-Abelian BPS monopoles in N=4 gauged supergravity}.
\newblock {\em Phys.Rev.Lett.}, {\bbf 79},\ 3343--3346, 1997.

\bibitem{Chamseddine98}
A.H. Chamseddine and M.S. Volkov.
\newblock {\ssl Non-Abelian solitons in N=4 gauged supergravity and leading
  order string theory}.
\newblock {\em Phys.Rev.}, {\bbf D 57},\ 6242--6254, 1998.

\bibitem{Donets93}
E.E. Donets and D.V. Gal'tsov.
\newblock {\ssl Stringy sphalerons and non-Abelian black holes}.
\newblock {\em Phys.Lett.}, {\bbf B 302},\ 411--418, 1993.

\bibitem{Duff95}
M.J. Duff, R.R. Khuri, and J.X. Lu.
\newblock {\ssl String solitons}.
\newblock {\em Phys.Rep.}, {\bf 259},\ 213--326, 1995.

\bibitem{Freedman78}
D.Z. Freedman and J.~Schwarz.
\newblock {\ssl N=4 supergravity model with local SU(2)$\times$SU(2)
  invariance}.
\newblock {\em Nucl.Phys.}, {\bf B 137},\ 333--339, 1978.

\bibitem{Hull98}
C.M. Hull.
\newblock {\ssl Gravitational duality, branes and charges}.
\newblock {\em Nucl.Phys.}, {\bf B 509},\ 216--251, 1998.

\bibitem{Lavrelashvili93}
G.~Lavrelashvili and D.~Maison.
\newblock {\ssl Regular and black hole solutions of Einstein -- Yang -- Mills
  dilaton theory}.
\newblock {\em Nucl.Phys.}, {\bbf B 410},\ 407--422, 1993.

\bibitem{Maldacena97}
J.~Maldacena.
\newblock {\ssl The large N limit of superconformal field theories and
  supergravity}.
\newblock {\em Adv.Theor.Math.Phys.}, {\bf 2},\ 231--252, 1998.

\bibitem{Strominger90}
A. Strominger.
\newblock {\ssl Heterotic solitons}.
\newblock {\em Nucl.Phys.}, {\bf B 343},\ 167--184, 1990.

\bibitem{Volkov98}
M.S. Volkov and D.V. Gal'tsov.
\newblock {\ssl Gravitating Non-Abelian solitons and black holes with
  Yang-Mills fields}.
\newblock To appear in {\em Physics Reports.}
\newblock hep-th/9810070.

\bibitem{Volkov99}
M.S.~Volkov, in preparation.

\end{thebibliography}

%\end{document}
%\newpage

\end{document}